# A Low Temperature Functioning CoFeB/MgO Based Perpendicular Magnetic Tunnel Junction for Cryogenic Nonvolatile Random Access Memory

*Lili Lang[1], Yujie Jiang[1], Fei Lu[1], Cailu Wang[1], Yizhang Chen[2], Andrew D. Kent[2], Li Ye[1]\**

[1] *Center for Excellence in Superconducting Electronics, Shanghai Institute of Microsystem and Information Technology, Chinese Academy of Sciences, Shanghai 200050, China*

[2] *Center for Quantum Phenomena, Department of Physics, New York University, New York, New York 10003, USA*

*e-mail: li_ye@mail.sim.ac.cn

**We investigated the low temperature performance of CoFeB/MgO based perpendicular magnetic tunnel junctions (pMTJs) by characterizing their quasi-static switching voltage, high speed pulse write error rate and endurance down to 9 K. pMTJ devices exhibited high magnetoresistance (>120%) and reliable (error rate<$10^{-4}$) bi-directional switching with 2 to 200 ns voltage pulses. The endurance of the devices at 9 K surpassed that at 300 K by three orders of magnitude under the same write conditions, functioning for more than $10^{12}$ cycles with 10 ns write pulses. The critical switching voltage at 9 K was observed to increase by 33% to 93%, depending on pulse duration, compared to that at 350 K. Ferromagnetic resonance and magnetization measurements on blanket pMTJ film stacks suggest that the increased switching voltage is associated with an increase in effective magnetic anisotropy and magnetization of free layer with decreasing temperature.**

**Our work demonstrates that CoFeB/MgO based pMTJs have great potential to enable cryogenic MRAM and that their low temperature magnetization and effective magnetic anisotropy can be further optimized to lower operating power and improve endurance.**

Magnetic random access memory (MRAM) is one of the most promising candidates for the next generation nonvolatile memory[1-3]. After years of development, embedded and standalone MRAM has been in manufacturing for multiple applications in a temperature range from 233 to 358 K, the typical commercial or industrial use operating range [4-6]. Recently, MRAM has been recognized to have potential to serve as cache or main working memory for a leading edge superconducting computer architecture[7], owing to its high density, fast operation, and compatibility with superconducting logic[9, 10]. However, recent research on various magnetic devices has yet to yield a practical cryogenic memory device. Spin orbit torque (or spin Hall effect, SHE) MRAM has shown fast switching at room temperature[11], but SHE MTJ-nTron bitcell scheme faces challenges of slow operation and large size [12]. Orthogonal spin transfer spin valve devices only achieve low error rate writing within a very limited pulse conditions at 4 K[13] and the read speed would be too slow for a working memory because of their low magnetoresistance. IBM proposed another memory bit cell integrating a field-switched in-plane magnetized MTJ and a Josephson junction[14]. Although this design is compatible with superconducting logic processors and expected to have low power consumption, it lacks low temperature experimental verification and faces scalability challenge as it requires additional wires to produce the switching fields.

Among various MRAM types, perpendicular spin transfer torque (STT)-MRAM is by far the most mature solution for a variety of applications. pMTJs could be the best candidate for cryogenic memory should it demonstrate appropriate functionality at cryogenic temperatures. Low temperature may bring some benefits in terms of data retention and read performance[15], but it also results in unfavorable impact on high speed writing: reduced thermal fluctuation decreases the initial STT[16, 17] and may increase the incubation delay[18]. Thus far reliable nanosecond pulse switching in pMTJ has not yet been experimentally observed below 10 K. Another key specification for a working memory is endurance: higher write voltage increases the stress on the pMTJ tunnel barrier and breakdown may occur in a shorter operation time. In brief, the key issues for application of pMTJ at low temperature lies in whether it can achieve appropriate comprehensive performance characteristics, e.g. large TMR, high speed switching, long endurance and low write error rate (WER). Their temperature dependence and underlying materials and device physics also need to be better understood to optimize pMTJs for cryogenic MRAM.

In this letter, we evaluated cryogenic characteristics of conventional CoFeB-MgO based pMTJ devices. We observed reliable low error rate ($<10^{-4}$) switching at temperatures as low as 9 K with pulse durations from 2 to 200 ns. Moreover, pMTJ could endure over $10^{12}$ cycles of 0.85 V amplitude, 10 ns write pulses at 9 K. From 300 to 9 K, the endurance improves by three orders of magnitude for the same stress conditions. This work demonstrates a functioning CoFeB/MgO based pMTJ device that is practical for use in cryogenic MRAM. The free layer magnetization $M_s$ and effective magnetic anisotropy $H_{Keff}$ are pointed out as the main attributes that determine the pMTJ's cryogenic switching characteristics.

The pMTJ stacks were shown in Fig. 1(a). The synthetic antiferromagnetic structure was used to pin the CoFeB reference layer and cancel the offset field. All film stacks were deposited on thermally oxidized Si substrates by dc and rf magnetron sputtering and then annealed in vacuum. The pMTJ devices with 80 nm-diameter-circle-shaped cross section were fabricated using electron beam lithography and Ar ion-milling. All electrical measurements were carried out in a close cycle Lakeshore cryogenics 40 GHz bandwidth probe station with a sample holder customized to accurately determine the sample temperature. All data shown in this work are the average of at least five different devices of a batch. No external magnetic field was applied in measurements of the switching characteristic.

Figure 1(b) shows a typical pMTJ resistance versus voltage ($V_{DC}$) at different temperatures. The parallel state resistance $R_P$ (~720 Ω) is nearly unchanged with temperature, while the antiparallel state resistance ($R_{AP}$) increases as the temperature goes down. This results in a nearly linear temperature dependence of the TMR (at 20 mV bias), as shown in Fig. 1(c). The magnetoresistance increases significantly at low temperature, by a factor of 1.8. With optimized pMTJ materials[19], over 300% magnetoresistance at low temperature could be achieved. It appears that $R_{AP}$ has a sharper dependence on voltage bias at low temperature within the range of ±100 mV. This could be explained by magnon assisted tunneling[20] or local heating effects at higher voltage bias. The switching voltages of AP-to-P ($V_{AP}$) and P-to-AP ($|V_P|$) direction both increase monotonically with decreasing temperature, and $|V_{AP}|$ is larger than $|V_P|$ at a given temperature, reflecting the asymmetry of STT[9], as shown in Fig. 1(d). $R_{AP}$ and $R_P$ at the switching voltage decreases just slightly as the temperature decreases from 350 K to 9 K, as exhibited in Fig. 1(e). This small change can be attributed to the counteracting effects of the lower $R_{AP(P)}$ at higher $V_{DC}$ for a fixed temperature and the higher $R_{AP(P)}$ at lower temperatures with the same $V_{DC}$. Therefore, the increase of $V_{AP(P)}$ as temperature decreases is mainly caused by the increase in the switching current $I_{AP(P)}$.

Nanosecond pulse switching measurements were carried out to evaluate write speed and reliability of pMTJs from 350 to 9 K. In the measurement, the voltage pulse signals consisted of reset and set square voltage pulses with opposite polarity. The reset pulse at a fixed width of 200 ns was applied to initialize the pMTJs to a known resistance state, and then a trigger applied the set pulse to the device after 500 μs. If the set pulse lead to free layer switching, the oscilloscope captured the event. The aforementioned procedure was repeated $10^4$ times for each set pulse amplitude and pulse width.

Figure 2(a)-(c) show the WER (equal to the probability of not switching) as function of pulse amplitude for several pulse durations. The 50% switching probability defines the critical switching voltage for pulse durations of 2, 10 and 200 ns. It is noteworthy that all magnetization switching occurs in the AP-to-P direction even at 9 K with a 2 ns pulse and the WER for P-to-AP switching can be ~$10^{-2}$. The critical

switching voltages both clearly increase with decreasing temperature down to 100 K and gradually saturate at 9 K. Moreover, the difference of switching voltage between AP-to-P and P-to-AP directions in a cryogenic environment is about 0.1 V, larger than that at 350 K (~0.03 V) for a given pulse width, which is consistent with quasi-static measurements. We also observed that the temperature has little influence on the slope of the WER versus pulse amplitude, in which the WER slope for AP-to-P switching is steeper than the other direction at a fixed pulse width. Figure 2(d)-(g) show WER diagrams of a typical pMTJ device for AP-to-P and P-to-AP direction at 9 K and 350 K, respectively. Switching voltage increases rapidly below 10 nanoseconds. For 2 ns duration pulses at 350 K, amplitudes of 1.55 V are needed to achieve a low WER (≤10$^{-2}$), whereas at 9 K ~2.0 V pulse amplitudes are needed, resulting in a writing energy at 9 K of about ten pJ.

Although switching voltage significantly increases at low temperature, the device resistance at the switching threshold barely changes (Fig. 1(d) and (e)). Therefore, the following analysis focuses on underlying physics that determines the switching current. The static and dynamic magnetization properties of free layer play essential roles in the critical current I$_C$ based on the equation[21, 22]

$$I_c(T) = \frac{\alpha(T) e \gamma M_s(T) H_{keff}(T) V_{free}}{\mu_B g_{STT}}, \quad (1)$$

where α, γ and $V_{free}$ represent the Gilbert damping constant, gyromagnetic ratio and volume of free layer. $\mu_B$ is the Bohr magneton and $g_{STT}$ is STT efficiency related to spin polarization and the magnetization angle between free and reference layer[6, 23].

Here, a blanket film with the same stack as the pMTJ devices was used to investigate the magnetization properties of free layer. The free layer magnetization was measured using vibrating sample magnetometer (VSM) in the range from 350 to 9 K, as depicted in Figure 3(a). α and $H_{keff}$ of the free layer were obtained by ferromagnetic resonance (FMR) measurements: the film was placed on a coplanar waveguide and a perpendicular applied field was swept for microwave frequencies (f) in the range 22 to 32 GHz. The resonance field ($H_{res}$) and the full width at half-maximum (ΔH) were extracted by fitting a Lorentzian to the resonance absorption line. Figure 3(b) shows the $H_{keff}$ of free layer after fitting to the Kittel equation[24, 25]

$$f = \frac{\gamma \mu_0}{2\pi}(H_{res} + H_{keff}) = \frac{\gamma}{2\pi}\left(\mu_0 H_{res} - \delta N \mu_0 M_s + \frac{2K_1}{M_s}\right), \quad (2)$$

where $\mu_0$, $\delta N = N_z - N_x$ and $K_1$ are the permeability in free space, the difference in demagnetization factor, the first-order uniaxial perpendicular magnetic anisotropy, respectively. Considering that the demagnetization factor in the blanket film and pMTJ device, i.e. $\delta N$ was 1 for the blanket film, while $\delta N$ was 0.9374 for the pMTJ device with a diameter of 80 nm and free layer thickness of 2.2 nm[25]. The $H_{keff}$ of free layer in a pMTJ device extrapolated from Eq. (2) increases significantly from 0.24 T to 0.43T with decreasing temperature (Figure 3(b)). Furthermore, γ is independent of temperature (~187.5 GHz/T)[26] from the slope of $H_{res}$ versus f (Fig. 3(c)). The damping, α, of free layer in the film was calculated from a linear fit of ΔH versus f using the following formula[27]

$$\mu_0 \Delta H = \mu_0 \Delta H_0 + \frac{4\pi \alpha f}{\gamma}, \quad (3)$$

where $\Delta H_0$ is the inhomogeneous linewidth broadening. α[28, 29] is 5×10⁻³ and nearly independent of temperature (Fig. 3(c)).

Assuming the pMTJ fabrication process does not change the magnetic properties of the film and the damping, we can use the film properties to analyze the device switching current. $I_C$ is proportional to the product of $M_s$ and $H_{keff}$ for fixed $g_{STT}$ (Eq. (1)). Figure 3(d) shows the ratio of $\mu_0 M_s H_{keff}$ and $V_C$ for quasi-static switching at different temperatures normalized to that at 350 K, denoted as $\Gamma(T) = \mu_0 M_s(T) H_{keff}(T) / \mu_0 M_s(350K) H_{keff}(350K)$ and $\eta(T) = V_c(T)/V_c(350K)$, respectively. It is noted that the increase of $\Gamma$ versus temperature (green solid line) follows a similar trend to η (green rhombi), which implies that $\eta(T)$ is likely associated with the change free layer's magnetic characteristics. $\Gamma$ deviates from η at low temperature. This may arise from the slight decrease of $R_{AP(P)@write}$ (Fig. 1(e)) or an enhancement of $g_{STT}(T)$ at lower temperatures. η for three different pulse widths is also presented in Fig. 3(d) for comparison. The slope of η versus temperature decreases as the pulse width decreases from 200 to 2 ns, although the amplitude of $V_C$ in the ballistic regime is ~4 times higher than that in thermally activated switching regime (Fig. 2(d)-(g)). Especially, from 350 to 9 K, the η at 2 ns increases by 32.5%, whereas η for quasi-static measurement increases by a factor of 1.93. Switching for long pulse durations is more affected by thermal fluctuations and therefore more sensitive to temperature changes. A similar trend was observed for low WER (<10$^{-2}$) switching voltage.

The endurance measurement was similar to the WER measurement. The reset and set pulses were applied with 5ns interval time. The voltage pulses were continuously triggered until the MgO barrier layer was damaged. Damage is associated with a sudden drop of resistance below 200 Ω. If no damage occurs, given time limitations, the test is stopped after $10^{11}$ stress cycles at a given condition (blue dashed line in Figure 4). All endurance tests are done using a pulse width of 10 ns. The cycle, at which 60% of pMTJ devices endure 10 ns write pulses, is defined as $N_W$. Figure 4 shows $N_W$ versus $V_{set}$ at 9 and 300 K. Remarkably, $N_W$ at 9 K is three orders of magnitude larger than that at 300 K for $V_{set}$=0.94 V, indicating that the lifetime of pMTJs is expected to become much longer at lower temperatures. This observation is consistent with the expectation that electromigration is reduced at low temperature. Furthermore, ln($N_W$) seems to depend linearly on voltage at 9 K and 300 K: the $N_W$ is estimated to improve by one order of magnitude when $V_{set}$ decreases by 47 mV at 9 K and 38 mV at 300 K. The $N_W$ at normal operation condition (~0.85 V at 9 K, ~0.66 V at 300 K) extrapolates to over $10^{12}$, assuming a continued linear dependence on pulse amplitude. Further development can focus on reducing switching voltage and improving endurance by optimizing the low temperature magnetic properties of the free layer.

In summary, the cryogenic switching characteristics of CoFeB/MgO based pMTJs were explored by quasi-static, WER and endurance measurement in a temperature range of from 9 to 350 K. It is demonstrated that pMTJ devices are suitable component for a cryogenic memory bitcell, showing high speed reliable switching even at 2 ns and endurance over $10^{12}$ cycles at 0.85 V and 10 ns write pulse width. The quasi-static $V_C$ monotonically increased, doubling with decreasing temperature down to 9 K, mainly resulting from the change of $I_C$ associated with an increase of $M_s H_{keff}$ at low temperatures. Moreover, the Endurance at 9 K improves by three orders of magnitude comparing to 300 K under the same stress conditions. Our results are also helpful to reducing the write energy by optimizing the stack structure of the free layer.

Note added: We are aware of a related research by Dr. Andrew D. Kent at New York University. Their work focuses on the low-temperature switching characteristics of pMTJs at ballistic regime. Their paper is also being posted on the arXivs.

This work was supported by the Strategic Priority Research Program of CAS under Grant XDA18000000. The FMR measurements at NYU were supported by NSF-DMR-1610416.

# Reference

[1]S. Tehrani, J. M. Slaughter, E. Chen, M. Durlam, J. Shi, and M. DeHerrera, IEEE Trans. Magn., **35**, 2814 (1999).

[2]N. Nishimura, T. Hirai, A. Koganei, T. Ikeda, K. Okano, Y. Sekiguchi, and Y. Osada, J. Appl. Phys., **91**, 5246 (2002).

[3]S. S. P. Parkin, K. P. Roche, M. G. Samant, P. M. Rice, R. B. Beyers, R. E. Scheuerlein, E. J. O'Sullivan, S. L. Brown, J. Bucchigano, D. W. Abraham, Y. Lu, M. Rooks, P. L. Trouilloud, R. A. Wanner, and W. J. Gallagher, J. Appl. Phys., **85**, 5828 (1999).

[4]C. J. Lin, S. H. Kang, Y. J. Wang, K. Lee, X. Zhu, W. C. Chen, X. Li, W. N. Hsu, Y. C. Kao, M. T. Liu, W. C. Chen, Y. Lin, M. Nowak, N. Yu, and L. Tran, IEEE Int. Electron Devices Meet., 1 (2009).

[5]X. Bi, H. Li, and X. Wang, IEEE Trans. Magn., **48**, 3821 (2012).

[6]Y. Wang, H. Cai, L. A. B. Naviner, Y. Zhang, J. O. Klein, and W. S. Zhao, Microelectron. Reliab., **55**, 1649 (2015).

[7]D. S. Holmes, A. L. Ripple, and M. A. Manheimer, IEEE Trans. Appl. Supercond., **23**, 1701610 (2013).

[8]W. J. Gallagher, and S. S. P. Parkin, IBM Journal of Research and Development, **50**, 5 (2006).

[9]L. Rehm, V. Sluka, G. E. Rowlands, M. H. Nguyen, T. A. Ohki, and A. D. Kent, Appl. Phys. Lett., **114**, 212402 (2019).

[10]G. E. Rowlands, S. V. Aradhya, S. Shi, E. H. Yandel, J. Oh, D. C. Ralph, and R. A. Buhrman, Appl. Phys. Lett., **110**, 122402 (2017).

[11]S. V. Aradhya, G. E. Rowlands, J. Oh, D. C. Ralph, and R. A. Buhrman, Nano Lett., **16**, 5987 (2016).

[12]M.-H. Nguyen, G. J. Ribeill, M. Gustafsson, S. Shi, S. V. Aradhya, A. P. Wagner, L. M. Ranzani, L. Zhu, R. Baghdadi, B. Butters, E. Toomey, M. Colangelo, P. A. Truitt, A. J. Salim, D. McAllister, D. Yohannes, S. R. Cheng, R. Lazarus, O. Mukhanov, K. K. Berggren, R. A. Buhrman, G. E. Rowlands, and T. A. Ohki, arXiv preprint arXiv: **1907.00942** (2019).

[13]G. E. Rowlands, C. A. Ryan, L. Ye, L. Rehm, D. Pinna, A. D. Kent, and T. A. Ohki, Sci. Rep., **9**, 803 (2019).

[14]J. -B. Yau, Y. -K. -K. Fung, and G. W. Gibson, IEEE International Conference on Rebooting Computing, **1** (2017).

[15]K. Cao, H. Li, W. Cai, J. Wei, L. Wang, Y. Hu, Q. Jiang, H. Cui, C. Zhao, and W. Zhao, IEEE Trans. Magn., **55**, 1 (2019).

[16]J. Kim, A. Chen, B. Behin-Aein, S. Kumar, J.-P. Wang, and C. H. Kim, IEEE custom integrated circuits conference, **1** (2015).

[17]M. Frankowski, M. Czapkiewicz, W. Skowroński, and T. Stobiecki, Phys. B, **435**, 105 (2014).

[18]T. Devolder, J. Hayakawa, K. Ito, H. Takahashi, S. Ikeda, P. Crozat, N. Zerounian, J. V. Kim, C. Chappert, and H. Ohno, Phys. Rev. Lett., **100**, 057206 (2008).

[19]Y.Huai, H. Gan, Z. Wang, P. Xu, X. Hao, B. K. Yen, R. Malmhall, N. Pakala, C. Wang, J. Zhang, Y. Zhou, D. Jung, K. Satoh, R. Wang, L. Xue, and M. Pakala, Appl. Phys. Lett., **112**, 092402 (2018).

[20]S. Zhang, P. M. Levy, A. C. Marley, and S. S. P. Parkin, Phys. Rev. Lett., **79**, 3744 (1997).

[21]V. B. Naik, H. Meng, and R. Sbiaa, AIP Adv., **2**, 042182 (2012).

[22]H. Liu, D. Bedau, J. Z. Sun, S. Mangin, E. E. Fullerton, J. A. Katine, and A. D. Kent, J. Magn. Magn. Mater., **358-359**, 233 (2014).

[23]Y. Wang, H. Cai, L. A. D. B. Naviner, Y. Zhang, X. Zhao, E. Deng, J.-O. Klein, and W. Zhao, IEEE Trans. on Electron Devices, **63**, 1762 (2016).

[24]J. M. Beaujour, D. Ravelosona, I. Tudosa, E. E. Fullerton, and A. D. Kent, Phys. Rev. B, **80**, 180415(R) (2009).

[25]K. Watanabe, B. Jinnai, S. Fukami, H. Sato, and H. Ohno, Nat. Commun., **9**, 663 (2018).

[26]S. Tacchi, R. E. Troncoso, M. Ahlberg, G. Gubbiotti, M. Madami, J. Akerman, and P. Landeros, Phys. Rev. Lett., **118**, 147201 (2017).

[27]M. P. R. Sabino, S. Ter Lim, and M. Tran, Appl. Phys. Express, **7**, 093002 (2014).

[28]X. Liu, W. Zhang, M. J. Carter, and G. Xiao, J. Appl. Phys., **110**, 033910 (2011).

[29]S. Iihama, Q. Ma, T. Kubota, S. Mizukami, Y. Ando, and T. Miyazaki, Appl. Phys. Express, **5**, 083001 (2012).

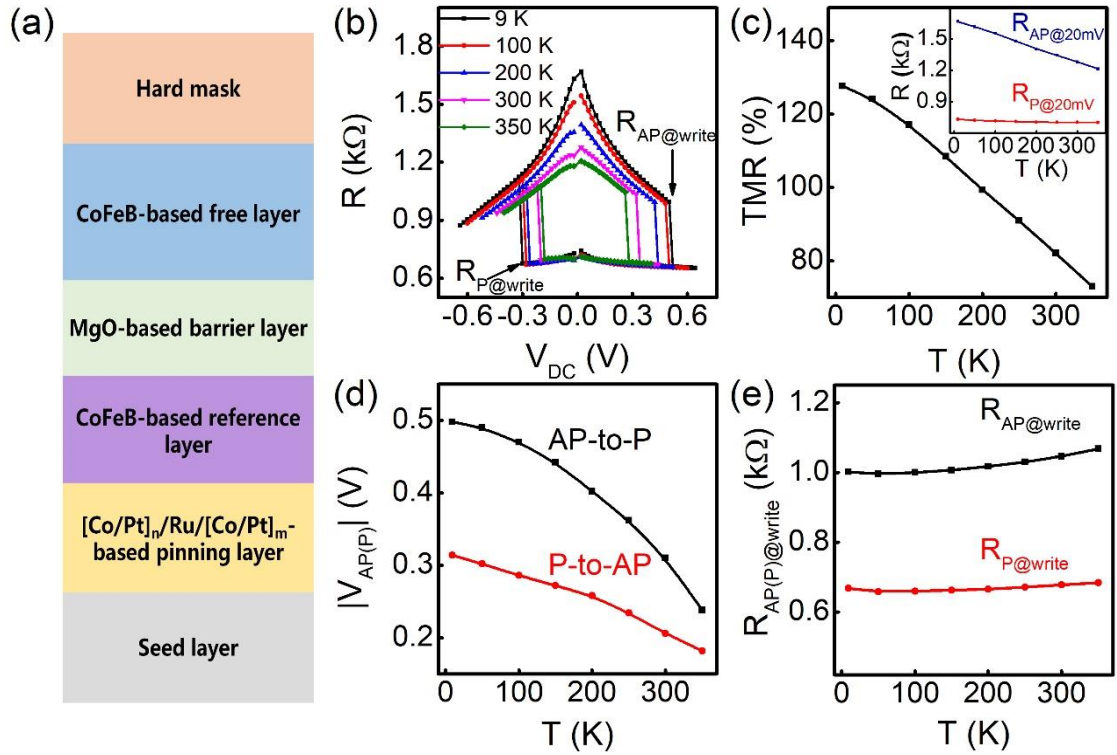

Fig.1 (a) Schematic of pMTJ stacks. (b) Resistance of a typical device as a function of $V_{DC}$ at different temperatures. (c), (d) and (e) show the TMR, $|V_{AP(P)}|$, $R_{AP@write}$ and $R_{P@write}$ as a function of temperatures, respectively. The insert in (b) shows the temperature dependencies of resistance for AP and P state resistances at a 20 mV bias voltage.

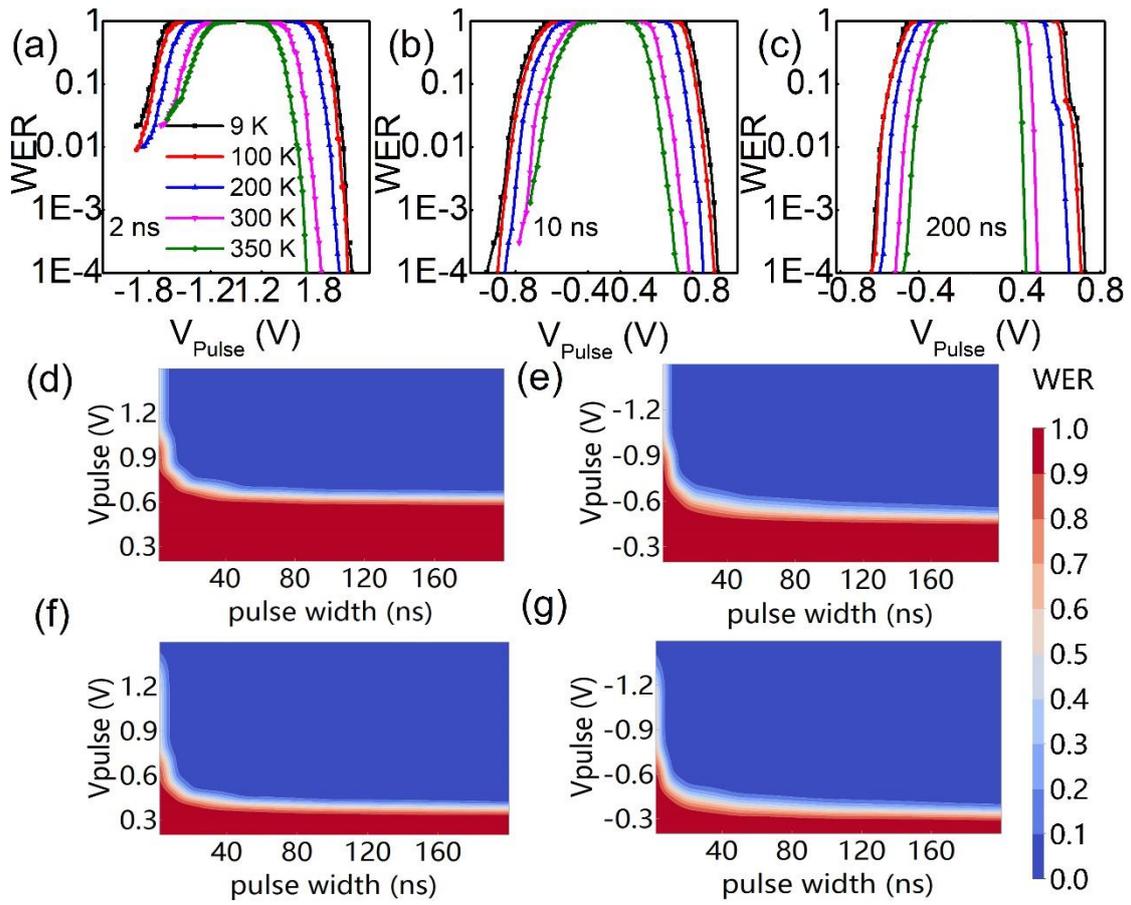

Fig. 2 The WER of a typical pMTJ device as a function of pulse voltage at 2 ns (a), 10 ns (b) and 200 ns (c). The WER diagrams for AP-to-P (d, f) and P-to-AP (e, g) at 9 K and 350 K, respectively.

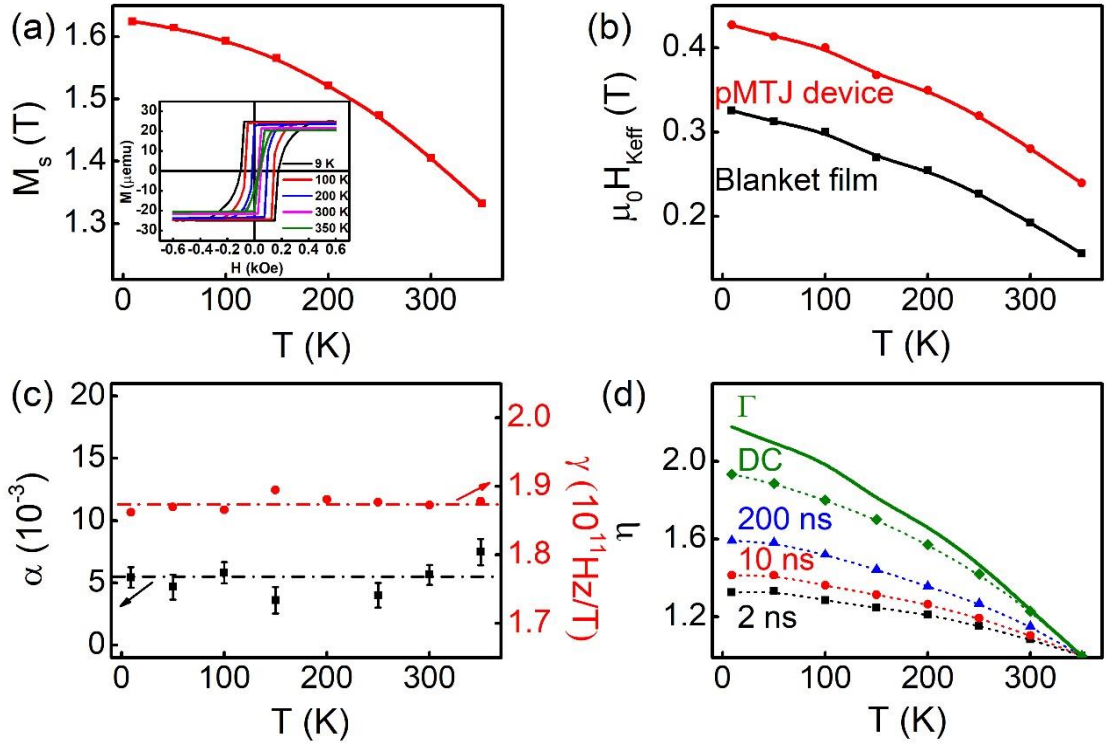

Fig. 3 (a) The $M_s$ of free layer, (b) $H_{keff}$ of free layer in the blanket film (black squares) and pMTJ device (red dots), (c) α (black squares) and γ (red dots) of free layer as a function of temperatures. (d) The Γ (solid green line) and η for quasi-static (green rhombi) and pulse switching (black squares at 2 ns, red dots at 10 ns and blue triangles at 200 ns) as a function of temperature. The inset in (a) shows the minor magnetic hysteresis loops of blanket film at different temperatures.

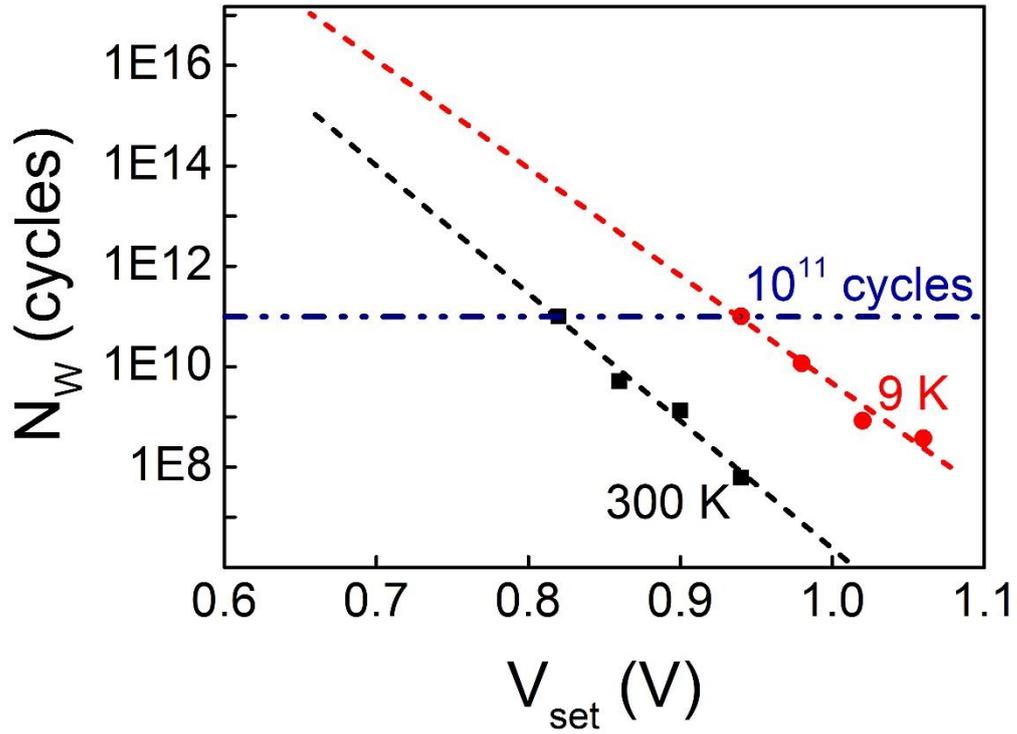

Fig.4 Endurance as a function of $V_{set}$ at 9 K and 300 K. The black and read dash lines are obtained from the linear fit of $lnN_W$ versus $V_{set}$. The horizontal dash line denotes the maximum pulse cycles in our endurance measurement.

16